\documentclass[11pt,twoside]{article}
\usepackage{asp2010}

\resetcounters

\begin{document}

\title{CO in GN20: The Nature of a z=4 Submillimeter Galaxy}
\author{J.~A.~Hodge,$^1$ C.~C.~Carilli,$^2$ F.~Walter,$^1$ W.~J.~G.~de~Blok,$^3$ D.~Riechers,$^4$ and E.~Daddi$^5$
\affil{$^1$MPIA, K\"{o}nigstuhl 17, 69117 Heidelberg, Germany}
\affil{$^2$NRAO, P.O. Box 0, Socorro, NM 87801-0387, USA}
\affil{$^3$ACGC, Astronomy Department, University of Cape Town, Private Bag X3, 7700 Rondebosch, Republic of South Africa}
\affil{$^4$Department of Astronomy, California Institute of Technology, MC 249-17, 1200 East California Boulevard, Pasadena, CA 91125, USA}
\affil{$^5$CEA, Laboratoire AIM-CNRS-Universit\'{e} Paris Diderot, Irfu/SAp, Orme des Merisiers, F-91191 Gif-sur-Yvette, France}}

\begin{abstract}
We present a study of the formation of clustered, massive galaxies at large look--back times, via high resolution spectroscopic imaging of CO in the unique GN20
proto-cluster. 
The data reveal evidence for rich structure and gas dynamics in unprecedented detail, 
allowing us to image the molecular gas with a resolution of only 1 kpc just 1.5 Gyr after the Big Bang.
These state-of-the-art data give new insight into the detailed physical processes involved in early massive galaxy formation, and they provide a first glimpse of the morphological studies that will become feasible on a regular basis with ALMA. 
\end{abstract}

\section{INTRODUCTION}
\label{Intro}

The majority of submillimeter galaxies \citep[SMGs;][]{2002PhR...369..111B} are believed to be starburst--dominated major mergers \citep[e.g.,][]{2003ApJ...599...92C}.
Indeed, there is direct evidence for multiple CO components and/or disturbed kinematics in some SMGs, supporting the merger picture \citep[e.g.,][]{2008ApJ...680..246T, 2010ApJ...724..233E, 2011MNRAS.412.1913I, 2011ApJ...739L..31R}. 
However, it has recently been suggested that other mechanisms that drive extreme star formation rates may also be at play. 
In particular, the cold mode accretion phenomenon \citep[CMA; e.g.][]{2009ApJ...703..785D, 2009Natur.457..451D} has been extended to SMGs by a number of authors \citep{2001astro.ph..7290F, 2006ApJ...639..672F, 2010MNRAS.404.1355D}. 
In this scenario, SMGs are massive galaxies sitting at the centers of deep potential wells and fed by smooth accretion.  

GN20 is an interesting SMG in this regard.
Situated at z$=$4.05, CO observations hint at a large spatial extent and ordered rotation \citep{2010ApJ...714.1407C}.
However, the previous VLA observations had some severe spectral limitations; 
in particular, the limited bandwidth truncated the line profiles, and the reliance on continuum mode precluded any information on kinematics.
We therefore obtained 124 hours on the Expanded Very Large Array (EVLA) to image the CO(2--1) emission in the GN20 field in the B-- and D--configurations. 
We centered the two 128 MHz IFs at 45.592 GHz and 45.720 GHz, for a total bandwidth of 246 MHz or 1600 km s$^{-1}$.  Each of the two IFs had 64 channels, corresponding to $\sim$13 km s$^{-1}$ per channel.  
The $\sim$20 hours of lower--resolution D--array data were presented in the EVLA Special Issue of ApJ by \citet{2011arXiv1105.2451C}. Here, we present the full dataset (B+D--configurations) on GN20, providing greatly improved spatial resolution and image fidelity.

\section{RESULTS \& ANALYSIS}
\label{results}

\begin{figure}[tbp]
\centering
\includegraphics[scale=0.5]{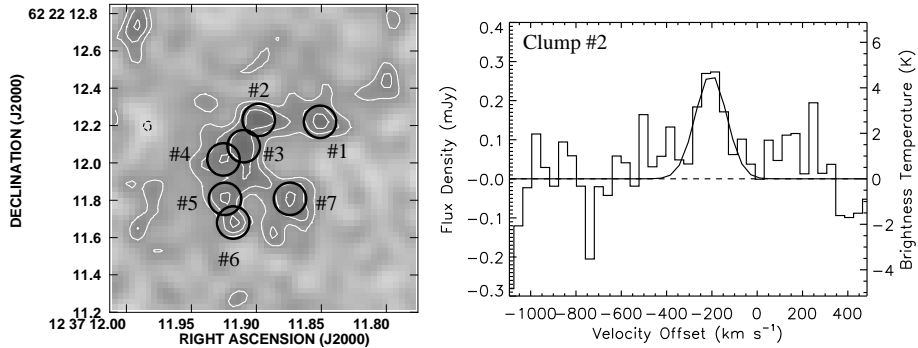}
\caption{Left: CO(2--1) velocity--averaged map over 780 km s$^{-1}$ at 0.19$^{\prime\prime}$ resolution. The circles identify molecular gas clumps and are the same size as the beam. The rms noise is 19 $\mu$Jy bm$^{-1}$, and contours are given in levels of 1$\sigma$ starting at $\pm$2$\sigma$. Right: Example spectrum for Clump \#2 at 39 km s$^{-1}$ resolution. Velocity offset is with respect to the systemic velocity of GN20.}
\label{fig:GN20_clumps}
\end{figure}

\subsection{Molecular Gas Clumps}
\label{clumps}

A velocity--averaged map is shown in Figure~\ref{fig:GN20_clumps} (left). 
The gas appears to be resolved into multiple star--forming clumps,
where we define as a clump anything with an H$_2$ column density greater than 6290 $\times$ ($\alpha/0.8$) M$_{\odot}$ pc$^{-2}$, or 
$\sim$35 times the average surface brightness of local giant molecular clouds \citep[GMCs;][]{1987ApJ...319..730S}.
Taking the clump size as the beamsize, this corresponds to a velocity--averaged CO flux density threshold of 76 $\mu$Jy beam$^{-1}$, or the 4$\sigma$ contour in the map. 
The clumps defined in this way have assumed sizes of 1.3 kpc (our beamsize), many orders of magnitude larger than even the giant molecular clouds in the nearby universe.

Using this definition of a molecular gas clump, we identified seven clumps in GN20 (identified by the circles in Figure~\ref{fig:GN20_clumps}).  
We then extracted a spectrum at the peak position of each clump.
While many of the clump spectra have low S/N, they allow us to estimate the properties of individual gas clumps for the first time.

Using Gaussian fits to the spectra, we find that the clumps have linewidths of $\sim$100--500 km s$^{-1}$ (FWHM)
and mass surface densities of 2,700--6,200 $\times$ ($\alpha/0.8$) M$_{\odot}$ pc$^{-2}$.
Derived brightness temperatures range from $\sim$11 K to 23 K (rest frame), approaching the dust temperature \citep[33K;][]{2011ApJ...740L..15M} and indicating that we are close to resolving the clumps. 
We find H$_2$ masses of 3.6--8.2 $\times$ 10$^{9}$ $\times$ $(\alpha/0.8)$ M$_{\sun}$. 
We therefore find that each clump makes up a few percent of the total mass of molecular gas.  
The combined mass of the clumps amounts to $\sim$30\% of the total gas mass.

\begin{figure}[tbp]
\centering
\includegraphics[scale=0.63]{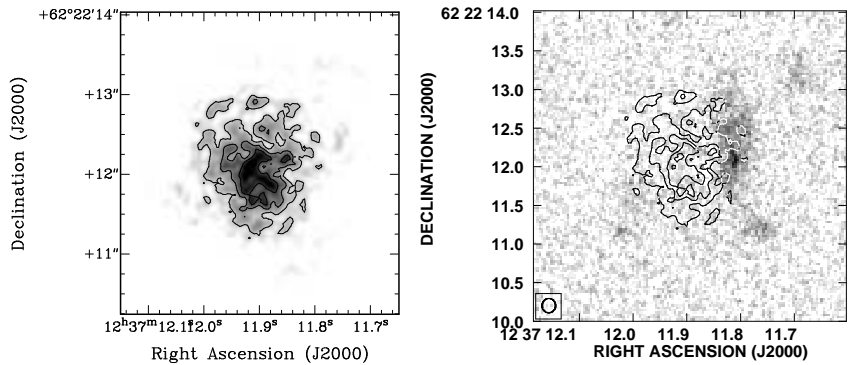}
\caption{Left: CO(2--1) 0th moment (i.e.\ integrated intensity) map at 0.19$^{\prime\prime}$ resolution. Contours start at (and are in steps of) 594 $\mu$Jy km s$^{-1}$. Right: 0th moment map overlaid on the HST$+$ACS z850--band image.}
\label{fig:GN20_mom0}
\end{figure}

\subsection{Dynamical Analysis}
\label{dyn}

\begin{figure}[tbp]
\centering
\includegraphics[scale=0.70]{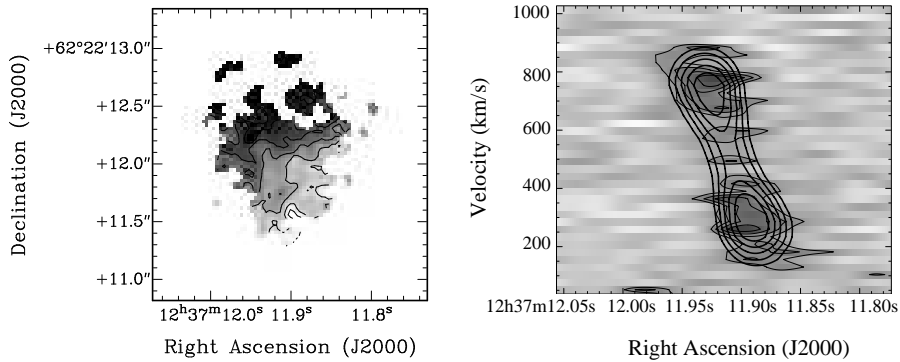}
\caption{Left: CO(2--1) 1st moment map at 0.19$^{\prime\prime}$ resolution. Contours show steps of 100 km s$^{-1}$, with the contour near the green representing the systemic velocity. Right: Major axis position--velocity diagram for CO(2--1), taken at a position angle of 25$^{\circ}$. The velocities on the vertical axis are relative.  Greyscale and thin countours show the observed data, and thick contours show the best fit model.}
\label{fig:GN20_mom1}
\end{figure}

For the dynamical modeling, we used the GALMOD task in the GIPSY package.
We carefully masked the data to create 0th and 1st moment maps for input to the modeling.
In the 0th moment map (Figure~\ref{fig:GN20_mom0}), we recover more diffuse emission than was seen in the original velocity--averaged map of Figure~\ref{fig:GN20_clumps}.
We estimate a radius for the disk of 1$^{\prime\prime}$ $\pm$ 0.3$^{\prime\prime}$, equivalent to ~$\sim$7 kpc ($\pm$ 2 kpc) at $z = 4.055$.
The more stringent masking applied to the velocity field is evident in the 1st moment map (Figure~\ref{fig:GN20_mom1}, left), which shows a zoomed--in region of the 0th moment map.
Although some noise is still present in the outskirts, a clear velocity gradient is apparent across the disk.

By comparing different models to the data, we find that the best--fit model is a rotating disk with an inclination of $i =$ 30$^{\circ}$ $\pm$ 15$^{\circ}$, 
a maximum rotational velocity of $v_{max}$ $=$ 575 $\pm$ 100 km s$^{-1}$, and a dispersion of $\delta = 100$ $\pm$ 30 km s$^{-1}$ (Figure~\ref{fig:GN20_mom1}, right).
While the resolution of the data make it difficult to constrain the exact shape of the rotation curve, we find that the velocity field is fully consistent with a rotating disk with a steeply--rising rotation curve that quickly flattens.	
We derive a dynamical mass for GN20 of 5.4 $\pm$ 2.4 $\times$ 10$^{11}$ M$_{\sun}$. 
We will use this estimate to set limits on the CO--to--H$_{2}$ mass conversion factor in the following.

If we assume that GN20 is composed entirely of molecular gas, then we derive a conversion factor of 3.4 M$_{\sun}$ (K km s$^{-1}$ pc$^{2}$)$^{-1}$.
If we instead assume roughly equal parts gas and stars, we derive $\alpha < 1.7$ M$_{\sun}$ (K km s$^{-1}$ pc$^{2}$)$^{-1}$.  
Here we have ignored the contributions from dust and dark matter, which will lower the limit even further.
While a galactic conversion factor of $\sim$4.3 M$_{\sun}$ (K km s$^{-1}$ pc$^{2}$)$^{-1}$ may therefore still be possible (within the uncertainties), we consider this an extreme case.
Our constraints on $\alpha$ are consistent with the limit $\alpha$ $<$ 1.0 derived using the local $M_{gas}/M_{dust}$ vs. metallicity relation \citep{2011ApJ...740L..15M}.  

\section{DISCUSSION}
\label{discussion}

The high--quality, high--resolution data on GN20 presented in this work reveal an extended gas reservoir, 14 $\pm$ 4 kpc in diameter, which is consistent with a rotating disk.
The data further resolve the reservoir into multiple, kpc--sized gas clumps 
which each constitute a few percent of the total gas mass and together comprise only 30\% of the total gas mass.
These observations may favor a scenario where the star formation is fueled by a process other than a major merger, e.g., cold mode accretion.
To determine if this is in fact the case, a crucial question must be addressed: is it possible to have a highly--obscured starburst without a major merger? This is a question that, due to the faintness of this particular source, may have to await the launch of JWST.

\acknowledgements We thank Glenn Morrson, Desika Narayanan, and Benjamin Weiner for useful comments and discussions. 

\bibliography{Hodge.bib}

\begin{thebibliography}{}
\expandafter\ifx\csname natexlab\endcsname\relax\def\natexlab#1{#1}\fi
\expandafter\ifx\csname url\endcsname\relax
  \def\url#1{\texttt{#1}}\fi
\expandafter\ifx\csname urlprefix\endcsname\relax\def\urlprefix{URL }\fi
\providecommand{\eprint}[2][]{\url{#2}}

\bibitem[{{Blain} et~al.(2002)}]{2002PhR...369..111B}
{Blain}, A.~W., et~al. 2002, Physics Reports, 369, 111.
  \eprint{arXiv:astro-ph/0202228}

\bibitem[{{Carilli} et~al.(2010)}]{2010ApJ...714.1407C}
{Carilli}, C.~L., et~al. 2010, \apj, 714, 1407. \eprint{1002.3838}

\bibitem[{{Carilli} et~al.(2011)}]{2011arXiv1105.2451C}
--- 2011, ArXiv e-prints. \eprint{1105.2451}

\bibitem[{{Chapman} et~al.(2003)}]{2003ApJ...599...92C}
{Chapman}, S.~C., et~al. 2003, \apj, 599, 92. \eprint{arXiv:astro-ph/0308197}

\bibitem[{{Dav{\'e}} et~al.(2010)}]{2010MNRAS.404.1355D}
{Dav{\'e}}, R., et~al. 2010, \mnras, 404, 1355. \eprint{0909.4078}

\bibitem[{{Dekel} et~al.(2009{\natexlab{a}}){Dekel}, {Sari}, \&
  {Ceverino}}]{2009ApJ...703..785D}
{Dekel}, A., {Sari}, R., \& {Ceverino}, D. 2009{\natexlab{a}}, \apj, 703, 785.
  \eprint{0901.2458}

\bibitem[{{Dekel} et~al.(2009{\natexlab{b}})}]{2009Natur.457..451D}
{Dekel}, A., et~al. 2009{\natexlab{b}}, \nat, 457, 451. \eprint{0808.0553}

\bibitem[{{Engel} et~al.(2010)}]{2010ApJ...724..233E}
{Engel}, H., et~al. 2010, \apj, 724, 233

\bibitem[{{Fardal} et~al.(2001)}]{2001astro.ph..7290F}
{Fardal}, M.~A., et~al. 2001, ArXiv Astrophysics e-prints.
  \eprint{arXiv:astro-ph/0107290}

\bibitem[{{Finlator} et~al.(2006)}]{2006ApJ...639..672F}
{Finlator}, K., et~al. 2006, \apj, 639, 672. \eprint{arXiv:astro-ph/0507719}

\bibitem[{{Ivison} et~al.(2011)}]{2011MNRAS.412.1913I}
{Ivison}, R.~J., et~al. 2011, \mnras, 412, 1913. \eprint{1009.0749}

\bibitem[{{Magdis} et~al.(2011)}]{2011ApJ...740L..15M}
{Magdis}, G.~E., et~al. 2011, \apjl, 740, L15. \eprint{1109.1140}

\bibitem[{{Riechers} et~al.(2011)}]{2011ApJ...739L..31R}
{Riechers}, D.~A., et~al. 2011, \apjl, 739, L31. \eprint{1105.4177}

\bibitem[{{Solomon} et~al.(1987){Solomon}, {Rivolo}, {Barrett}, \&
  {Yahil}}]{1987ApJ...319..730S}
{Solomon}, P.~M., {Rivolo}, A.~R., {Barrett}, J., \& {Yahil}, A. 1987, \apj,
  319, 730

\bibitem[{{Tacconi} et~al.(2008)}]{2008ApJ...680..246T}
{Tacconi}, L.~J., et~al. 2008, \apj, 680, 246. \eprint{0801.3650}

\end{thebibliography}

\end{document}